\documentclass[letter]{aa} 
\usepackage{graphicx}
\usepackage{txfonts}
\def\kms{km\,s$^{-1}$}
\def\Ha{H$\alpha$}

\def\HeI{He\,{\sc i}}
\def\HeII{He\,{\sc ii}}
\def\NII{N\,{\sc ii}}
\def\OII{O\,{\sc ii}}
\def\OIII{O\,{\sc iii}}
\def\ergcms{erg cm$^{-2}$ s$^{-1}$}

\begin{document}
   \title{Asymmetries in the Type IIn SN~2010jl\thanks{Based on observations 
   collected at the German-Spanish  Astronomical Center, Calar Alto, jointly operated
   by the Max-Planck-Institut f\"ur Astronomie Heidelberg and the Instituto de Astrofisica
   de Andalucia (CSIC).}}
  \subtitle{}
   \author{F. Patat\inst{1}
   \and    
    S. Taubenberger\inst{2}
   \and
    S. Benetti\inst{3}
    \and 
    A. Pastorello \inst{4}
    \and
    A. Harutyunyan\inst{5}
}


   \institute{European Organization for Astronomical Research in the 
	Southern Hemisphere (ESO), Karl-Schwarzschild-Str.  2,
              85748, Garching b. M\"unchen, Germany
              \email{fpatat@eso.org}
              \and 
              Max-Planck-Institut f\"ur Astrophysik, Karl-Schwarzschild-Str. 1,
              85741, Garching b. M\"unchen, Germany
              \and
               Istituto Nazionale di Astrofisica, Osservatorio Astronomico
               di Padova, v. Osservatorio n.5, I-35122, Padua, Italy
              \and
	      Queen's University - Belfast, BT7 1NN, Northern Ireland, UK
              \and
              INAF, Telescopio Nazionale Galileo. Rambla Jos\'e Ana 
              Fern\'andez P\'erez 7, 38712 Bre\~na Baja, TF - Spain 
}

   \date{Received November, 2010; accepted ...}

\abstract{}{}{}{}{} 
 
  \abstract
   {}
   {We study possible signs of asymmetry in the luminous Type IIn SN~2010jl, 
    to obtain independent information on the explosion geometry.
   }
   {We obtained optical linear 
   spectropolarimetry of SN~2010jl two weeks after
   the discovery, in the spectral range 3700--8800 \AA.}
   {The object exhibits a continuum polarization at a very significant
     and almost constant level (1.7-2.0\%). Marked line
     depolarization is seen at the positions of the strongest emission
     features, like H$\alpha$ and H$\beta$. This implies that the line
     forming region is well above the photosphere. The continuum
     polarization level (1.7--2.0\%) indicates a substantial
     asphericity, of axial ratio $\leq$0.7. The almost
     complete depolarization seen at H$\alpha$ suggests a very low
     level of interstellar polarization ($\leq$0.3\%). This rules out
     the presence of relevant amounts of dust in the progenitor
     environment at the time of our observations. From a polarimetric
     point of view, SN~2010jl appears to be very similar to the two other
     well studied Type IIn SNe 1997eg and 1998S, establishing a strong link
     within this class of objects.}
   {}


\authorrunning{F. Patat et al.}
\titlerunning{Asymmetries in the Type II-n SN~2010jl}

   \maketitle
%

\section{\label{sec:intro}Introduction}

Type IIn Supernovae (SNe) are characterized by intense,
composite-profile emission lines, and by the absence of the broad
P-Cygni absorption troughs that are the distinguishing feature of
{\it classical} Type II SNe (Schlegel \cite{schlegel90}). The early
spectra of these objects are explained in terms of a strong
interaction between the fast expanding SN ejecta and a dense, slow
moving circumstellar medium, fed by a significant mass loss undergone
by the progenitor star (Chugai \cite{chugai97}). However, many
questions about these objects, the mass loss history of their
progenitors, the geometry of the explosion and the circum-stellar material (CSM)
remain unanswered. Even the explanation for the origin of the broad
emission seen at the base of the intermediate and narrow components
has been questioned (Chugai \cite{chugai01}). With its unique
capability of retrieving geometrical information, spectropolarimetry
constitutes and independent way of attacking these open issues.

The bright SN~2010jl was discovered on November 3.5, 2010 (Newton \&
Puckett \cite{newton}). A spectrum obtained on November 5 revealed it
was a Type IIn (Benetti et al. \cite{benetti}), with the typical,
multi-component emission lines of hydrogen, and weaker helium
features. Based on archival images of the explosion site, Smith et
al. (\cite{smith}) concluded that the progenitor of this luminous was
probably a massive star, with an initial mass larger than 30
M$_\odot$.  Soon after the discovery, we started an optical/near-IR
follow-up using a number of ground-based facilities. The results will
be presented and discussed elsewhere. In this letter we focus on the
spectropolarimetry of SN~2010jl obtained about two weeks after its
discovery, with the aim of detecting possible signs of asymmetries in
the ejecta and/or in the CSM that could provide
additional constraints on the geometry of the explosion environment.

\section{\label{sec:obs}Observations and data reduction}

We observed SN~2010jl on November 18.2 UT, i.e. 14.7 days after its
discovery, using the Calar Alto Faint Object Spectrograph (CAFOS),
mounted at the 2.2 m telescope in Calar Alto, Spain (Meisenheimer
\cite{cafos}). In this multi-mode instrument, equipped with a
2K$\times$2K SITe-1d CCD (24$\mu m$ pixels, 0.53 arcsec/pixel),
polarimetry is performed by introducing into the optical path a
Wollaston prism (18$^{\prime\prime}$ throw) and a super-achromatic
half-wave plate (HWP). To reduce some known instrumental problems (see
Patat \& Romaniello \cite{patat06}), we used 4 half-wave plate angles
(0, 22.5, 45 and 67.5 degrees). All spectra were obtained with the
low-resolution B200 grism coupled with a 1.0 arcsec slit, giving a
spectral range 3300-8900 \AA, a dispersion of $\sim$4.7 \AA\/
px$^{-1}$, and a FWHM resolution of 14.0 \AA\/ (640 km s$^{-1}$ at
H$\alpha$).  For each HWP angle we obtained two exposures of 20
minutes each. The resulting signal-to-noise ratio per pixel on the
single beam is $\sim$180 at 5000 \AA. Data were bias and flat-field
corrected, and wavelength calibrated using standard tasks within IRAF.
Stokes parameters ($Q$, $U$), linear polarization degree ($P$), and
position angle ($\theta$) were computed by means of specific routines
written by us. Finally, polarization bias correction and error
estimates were performed following the prescriptions described by
Patat \& Romaniello (\cite{patat06}). The HWP zero-point angle
chromatism was derived using the polarization standard star BD+59d389
($P(V)$=6.7$\pm$0.2\%, $\theta$=98.1 degrees; Schmidt et
al. \cite{schmidt}), observed on November 18.8 UT with the same
instrumental setup. Instrumental polarization was analyzed using the
unpolarized star HD~14069 ($P(V)$=0.02$\pm$0.02\%; Schmidt et
al. \cite{schmidt}), and found to vary very slowly as a function of
wavelength between 0.3\% and 0.4\%. Once this correction is applied to
the data of BD+59d389, the resulting linear polarization degree is
consistent with the published values to within 0.1\%. To increase the
signal-to-noise ratio, the final Stokes parameters were binned in
$\sim$51.2 \AA\/ wide bins (11 pixels). This results in a nominal
RMS error in the polarization of 0.1\% at 5000\AA, where the continuum
signal-to-noise ratio per resolution element reaches its maximum
(580). Flux calibration was achieved by observing a spectrophotometric
standard star with the polarimetric optics inserted. Wavelengths were
corrected to the rest frame adopting a recession velocity of 3214 km
s$^{-1}$ for the host galaxy (Koribalski et al. \cite{kori}).

\begin{figure}
\centerline{
\includegraphics[width=95mm]{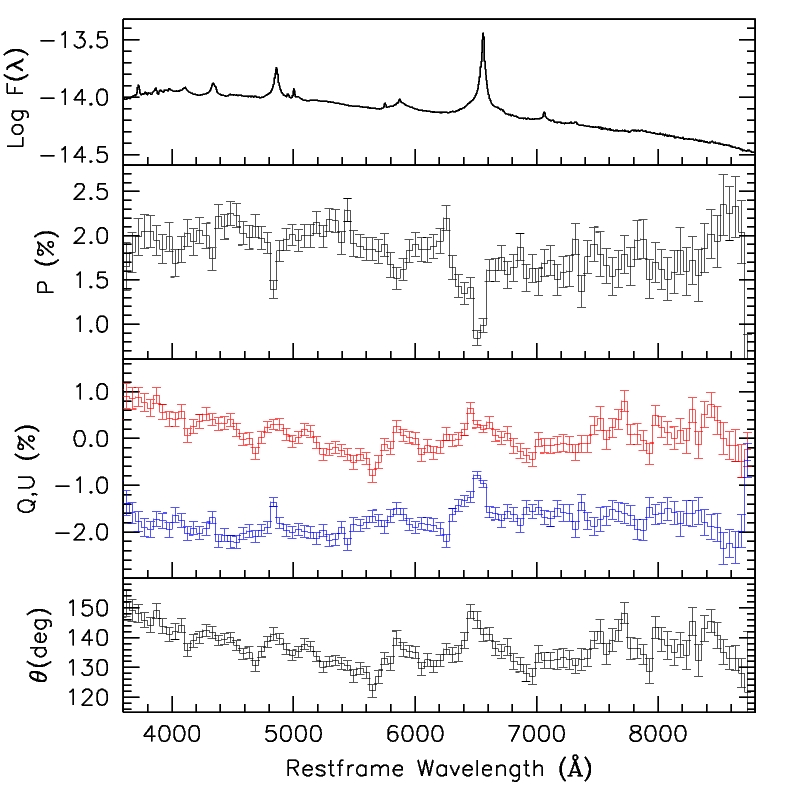}
}
\caption{\label{fig:pol}Spectropolarimetric data for SN~2010jl. From
  top to bottom: unbinned total flux spectrum ($\log F(\lambda)$ in erg
  s$^{-1}$ cm$^{-2}$ \AA$^{-1}$), linear polarization degree, Stokes
  parameters $Q$ (upper) and $U$ (lower), and polarization position
  angle on the plane of the sky.}
\end{figure}

\section{\label{sec:ext}Reddening and interstellar polarization}

The extinction arising within the Milky Way along the line of sight to
SN~2010jl is very small ($E_{B-V}$=0.027, Schlegel, Finkbeiner \&
Davis \cite{schlegel}). The inspection of a high resolution spectrum
($\lambda/\Delta\lambda\sim$46,000) obtained with the SARG Echelle
Spectrograph at the TNG on November 7, 2010 shows very weak
\ion{Na}{i} D lines at about zero velocity. The equivalent width (EW)
of the $D_2$ component is less than 100 m\AA, implying $E_{B-V}<$0.02
(Munari \& Zwitter \cite{munari}). Therefore, the limit on
interstellar polarization (ISP) within the Galaxy is $P_{MW}\leq$0.2\%
(Serkowski, Matheson \& Ford \cite{serkowski}; Whittet et
al. \cite{whittet}). The compilation of Heiles (\cite{heiles})
contains two stars within 3 degrees from the position of UGC~5189A.
Both have $P$=0.05\%, in full agreement with the low Galactic
extinction along the line of sight. Interstellar \ion{Na}{i} D lines
are detected in the high resolution spectrum at $\sim$3120 km
s$^{-1}$, possibly resulting from 2 or 3 separate components within 50
km s$^{-1}$. The total EW of the $D_2$ line is 0.20$\pm$0.02 \AA,
implying $E_{B-V}$=0.03 for a MW dust/gas mixture (Munari \& Zwitter
\cite{munari}). Although there is a mounting evidence that the
Serkowski law is not universal (Leonard \& Filippenko
\cite{leonard01}; Leonard et al. \cite{leonard02}; Maund et
al. \cite{maund07}; Patat et al. \cite{patat09,patat10}), we reckon it
is very plausible the ISP in the host galaxy is small, most likely
below 0.3\%. For this reason, in the following we consider the ISP
negligible. This assumption is independently bolstered by the lack of
reddening evidence in the flux spectrum (see Sect.~\ref{sec:flux}),
and the almost complete depolarization observed at the peak of
H$\alpha$ (see Sect.~\ref{sec:line}).

\section{\label{sec:flux}Flux spectrum}

The total flux spectrum of SN~2010jl shown in Fig.~\ref{fig:pol} (top
panel) is characterised by emission lines of H, He and CNO elements
superimposed on a blue continuum. This is nicely matched by a
blackbody of $\sim$\,7000 K after correcting for Galactic
reddening. With this correction there is no additional need for
host-galaxy reddening or multiple-temperature components (see the
discussion in Smith et al. \cite{smith}). This is fully in line with
the low reddening implied by the measured \ion{Na}{i} D EWs
(Sect.~\ref{sec:ext}). As often found in SNe IIn, a de-blending of the
\Ha\ emission reveals a broad component of $\sim$\,10,500 \kms\ FWHM,
with a blue-shift of $\sim$\,580 \kms, and an integrated flux of
$6.8\,\times\,10^{-13}$ \ergcms. An intermediate ($\sim$\,2400 \kms)
and an unresolved component ($\sim$\,640 \kms) are also present.  We
note that, based on higher resolution spectra, Smith et
al. (\cite{smith}) resolved the narrow component at 120 \kms\/ FWHM. A
similar width is measured in our TNG-SARG spectrum.  Apart from the
Balmer series of hydrogen, narrow unresolved emission lines of [\OII]
$\lambda3727$, [\OIII] $\lambda\lambda4959,5007$, [\NII]
$\lambda5755$, \HeI\ $\lambda5876$ and \HeI\ $\lambda7065$ are also
detected. For the \HeI\ lines, underlying broader components can be
discerned. There is also weak evidence of \HeII\ $\lambda4686$,
\HeI\ $\lambda7281$ and [\OII] $\lambda\lambda7320,7330$.  In general,
SN~2010jl is spectroscopically very similar to other well studied Type
IIn events, like SN~2007rt (Trundle et al. \cite{trundle} and references therein).

\section{\label{sec:specpol}Spectropolarimetry}

\subsection{\label{sec:cont}Continuum polarization}

The continuum appears to be polarized at a very significant level
across the whole spectral range (Fig.~\ref{fig:pol}). With the
remarkable exceptions of regions dominated by the main emission lines
(see next section), the polarization degree is roughly constant. The
mean level estimated on different portions of the continuum is:
2.00$\pm$0.04\% (3800--4600 \AA), 2.02$\pm$0.05\% (5000--5600 \AA),
and 1.67$\pm$0.03\% (6900--8400 \AA). On average, the polarization
is slightly higher in the blue than in the red. The analysis of
the $Q-U$ plane reveals that the data strongly cluster around one
single point, and there is no statistically significant correlation
between $Q$ and $U$. The averages within the spectral range 4000--8000
\AA\/ are 0.02$\pm$0.03\% and $-$1.77$\pm$0.03\% for $Q$ and $U$
respectively (the RMS deviation from the mean value is 0.3\% for
both parameters). For this reason, the position angle is almost
constant as a function of wavelength (see also Fig.~\ref{fig:pol},
bottom panel). The average values measured on the continuum in the
same ranges as above are 140.9$\pm$0.9, 132.1$\pm$1.0, and
136.0$\pm$0.9 degrees respectively (with RMS deviations of 3.5, 3.3,
and 5.2 degrees). The polarized signal is almost completely carried by
the $U$ component.

\subsection{\label{sec:line}Line polarization}

\begin{figure}
\centerline{
\includegraphics[width=90mm]{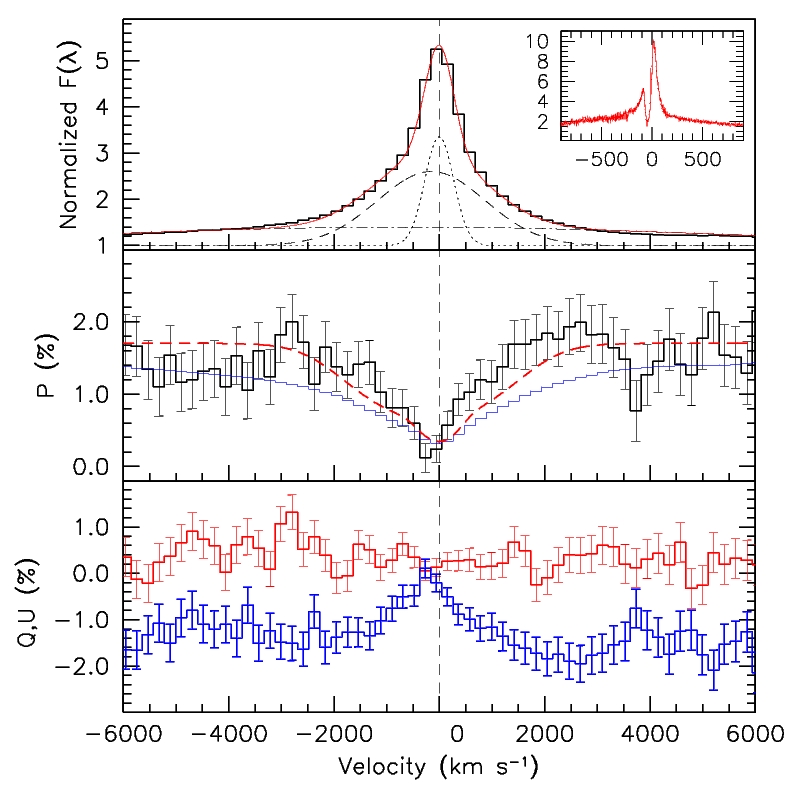}
}
\caption{\label{fig:ha}Spectropolarimetry of the H$\alpha$ line
  (unbinned data). {\bf Top panel}: continuum normalized flux profile. The thin
  curves trace the narrow (dotted), intermediate (dashed), and broad
  (dotted-dashed) components. The solid curve is the sum of the
  three. The inset shows the high-resolution SARG spectrum (see
  text). {\bf Mid panel}: polarization degree. The thin solid line traces the
  depolarization expected for dilution by the full line profile, while the
  dashed line is the same for the intermediate and narrow components
  only. {\bf Bottom panel}: Stokes parameters $Q$ (upper) and $U$
  (lower). }
\end{figure}

Significant depolarizations are observed at the positions of the main
emission features H$\alpha$, H$\beta$, H$\gamma$, and \ion{He}{i} 5876 \AA\/
(Fig.~\ref{fig:pol}). The level of depolarization appears to be
related to the line intensity, therefore strongly favoring continuum
dilution as the responsible mechanism. Given the marked enhancement in
the signal-to-noise across the H$\alpha$ emission, this line can be
studied at full spectral resolution (640 km s$^{-1}$). The
polarization profile is roughly symmetric, and $P$ departs from the
continuum level ($\sim$1.7\%) at about $\pm$3000 km s$^{-1}$
(i.e. where the intermediate component starts to dominate over the
broad one), to reach a minimum (0.12$\pm$0.20\%) around $-$250 km
s$^{-1}$, which is slightly blue-shifted with respect to the total
flux profile (Fig.~\ref{fig:ha}). Under the hypothesis that the
emission line is unpolarized, the polarization across the profile can be
computed using the semi-empirical relation proposed by McLean \&
Clarke (\cite{mclean}). The result is plotted in
Fig.~\ref{fig:ha} (mid panel, thin solid curve). Although the polarization
level at the center of the line is well reproduced, the wings of the
line display a residual polarization.  A much better match is obtained considering
only the intermediate and narrow components (mid panel, dashed curve). 
This suggests that the broad component is polarized to some extent
(see also Sect.~\ref{sec:disc}). All the variation takes place along
the $U$ component (Fig.~\ref{fig:ha}, bottom panel). As $Q$ is close
to 0, the points in the $Q-U$ plane move along the $U$ axis, so that
the position angle remains constant across the line profile. This is
what is expected for simple continuum depolarization dilution (see for
instance Hoffman et al. \cite{hoffman}). 

The presence of dust above the line forming region (e.g. in the circum- or inter-stellar
environment) would polarize the light irrespective of its
origin. i.e. affecting both the adjacent continuum and the
line. Admittedly, depending on its position angle, the ISP could also
act as a depolarizer. However, in order to have a null net
polarization, the component of the IS Stokes vector perpendicular to
the intrinsic polarization needs to have a fine tuned value, which
makes this possibility rather unlikely. We note that if this were the
case, then the intrinsic polarization would be higher than what is
observed. The almost complete depolarization seen at the center of
H$\alpha$ is strongly suggestive of a negligible ISP, thus confirming
the expectations based on reddening considerations
(Sect.~\ref{sec:ext}).

\section{\label{sec:disc}Discussion}

As Type IIn events are only 6-9\% of all core-collapse SNe (Li et
al. \cite{li10}), and the technique requires very high signal-to-noise
data, spectropolarimetry of this class of objects is rare (Wang \&
Wheeler \cite{wang08}). The only two published cases are those of
SNe~1997eg (Hoffman et al. \cite{hoffman}: 16, 44 and 93 days after
discovery), and 1998S (Leonard et al. \cite{leonard00}: 5 days after
discovery; Wang et al. \cite{wang01}: 28 and 59 days after discovery),
which reached a comparable absolute magnitude ($M_V\sim-$18.6),
which is significantly fainter than the peak magnitude reached by SN~2010jl
($M_V\sim-$20.6, Smith et al. \cite{smith}). Although the analysis was made difficult by the
uncertainty in the ISP correction, in the earliest epochs the two
objects displayed a significant and almost constant continuum
polarization (2.0-2.6\%), characterized by a constant position
angle. Additionally, marked line depolarizations were observed at the
positions of the most prominent emission lines (Leonard et
al. \cite{leonard00}; Wang et al. \cite{wang01}; Hoffman et
al.\cite{hoffman}). Our data show that SN~2010jl conforms strikingly
well to this behavior, hinting to a common geometry of the explosion
or, more in general, of the environment in which the explosion takes
place.

The basic physical facts that emerge from the early time data are:
{\it i)} the wavelength-independent continuum polarization; {\it ii)}
the constant position angle, and {\it iii}) the strong line
depolarization. These place SNe IIn in a very special niche within the
SN polarimetric zoo (see Wang \& Wheeler \cite{wang08} for a general
review). The most natural source of wavelength-independent linear
polarization in a stellar envelope is Thomson scattering by free
electrons. In this context, a non-null net polarization is indicative
of an asymmetry in the continuum-emitting region. In the case of an
axis-symmetric geometry, this would produce a wavelength-independent
 polarization angle (perpendicular to the major axis). Under the assumption of a
spheroidal electron-scattering atmosphere, the level of polarization
measured in SN~2010jl ($\sim$2\%) indicates a substantial asphericity,
of axial ratio $\leq$0.7 (H\"oflich \cite{hoeflich}). This value is
very similar to those derived for the other two known cases (Leonard
et al. \cite{leonard00}; Hoffman et al. \cite{hoffman}). As first
proposed by Leonard et al. (\cite{leonard00}), the simplest
explanation for the line depolarization in Type IIn SNe is the dilution of
polarized continuum light by unpolarized line recombination emission,
which has to take place well above the photosphere, in a region where
electron scattering becomes negligible. This picture was made more
complicated by the detection of a different behavior in the narrow
component, suggesting a different geometry for the regions where these
lines form (Leonard et al. \cite{leonard00}; Hoffman et
al. \cite{hoffman}). As the narrow component is not resolved in our
spectra, we cannot address this issue. However, we note that while the
core of the H$\alpha$ profile (corresponding to the narrow and intermediate 
components) is depolarized, this is not the case for the wings at velocities larger 
than 3000 \kms\/ (see Sect.~\ref{sec:line}). 

This might be understood in the scenario proposed by Chugai
(\cite{chugai01}), where the narrow component is essentially
unpolarized, while the broad wings are the result of electron
scattering. In this picture, the broad component does not arise in the
reverse-shocked ejecta (as commonly assumed), but rather in a dense,
fully ionized CSM shell dominated by electron scattering, where the
photosphere would also reside. A closer inspection of the binned
H$\alpha$ profile indeed shows a polarization bump (mostly in the $Q$
parameter), spanning the full breadth of the broad component
(Fig.~\ref{fig:pol}), and peaking at $\sim$0.4$\pm$0.1\% above the
adjacent continuum polarization level. The change of $\theta$ seen
across the bump suggests a different geometry for the photosphere and
the broad line forming region. Also, there is an excess of
polarization on the blue wing, similar to what was seen in SN~2007eg.
This possibly indicates that the receding side of the scattering
region is obscured either by its approaching side or by the ejecta (Hoffman
et al. \cite{hoffman}).

In a scenario where continuum polarization is acquired within the CSM, 
spectropolarimetry can hardly provide information on the asymmetry of the 
explosion itself, but only on the geometry of the circum-stellar environment. As
a consequence, the high level of
continuum polarization observed in Type IIn SNe would be interpreted as a
strong deviation from sphericity in the CSM and, in turn, in the
outflow of material during the mass loss episodes preceding the
explosion. This conundrum can be resolved only via multi-epoch, higher
spectral resolution spectropolarimetry, which with 8m-class telescopes
should be possible, at least for the most luminous events. This will
allow one to distinguish the different behavior of the various
components, and to study their evolution with time, as the ejecta
proceed into the CSM.

One important fact implied by our observations is that the low
polarization level measured at the peak of H$\alpha$ ($\leq$0.2\%) is
inconsistent with the presence of substantial amounts of dust along
the line of sight, including the immediate surroundings of the
explosion site. Therefore, although a significant quantity of CSM must
be present at the epoch of our observations (as revealed by the strong
signs of interaction), this material has to have a very low dust
content. Whether this is intrinsic or caused by dust evaporation by
the UV radiation cannot be concluded based on the available data.

\begin{acknowledgements}
We thank Calar Alto Observatory for allocation of Director's Discretionary Time to this project. This work has been conducted in the framework of the European collaboration "SN Variety and Nucleosynthesis Yields". \end{acknowledgements}


\begin{thebibliography}{}
\bibitem[2010]{benetti} Benetti, S., et al., 2010, CBET, 2536, 1
\bibitem[1997]{chugai97} Chugai, N.N., 1997, Ap\&SS, 252, 225
\bibitem[2001]{chugai01} Chugai, N.N., 2001, MNRAS, 326, 1448
\bibitem[1999]{heiles} Heiles, C., 1999, AJ, 119, 923
\bibitem[1991]{hoeflich} H\"oflich, P., 1991, A\&A, 246, 481
\bibitem[2008]{hoffman} Hoffman, J.L., et al., 2008, ApJ, 688, 1186
\bibitem[2004]{kori} Koribalski, B.S., et al., 2004, AJ 128, 16   
\bibitem[2000]{leonard00} Leonard, D.C., Filippenko, A.V., Aaron, J.B. 
        \& Matheson, T., 2000, ApJ, 536, 239
\bibitem[2001]{leonard01}Leonard, D.C. \& Filippenko, A.V., 2001,
	PASP, 113, 920
\bibitem[2002]{leonard02}Leonard, D.C., Filippenko, A.V., Chornock, R. \&
	Weidong, L., 2002, ApJ, 124, 2506
\bibitem[2010]{li10} Li, W., et al., 2010, MNRAS, submitted (arXiv:1006.4612)
\bibitem[2007]{maund07}Maund, J.R., Wheeler, J.C., Patat, et al.,
  2007, MNRAS, 381, 201
\bibitem[1979]{mclean} McLean, I.S. \& Clarke, D., 1979, MNRAS, 186, 245
\bibitem[1998]{cafos} Meisenheimer, K., 1998, User Guide to the CAFOS 2.2
\bibitem[1997]{munari} Munari, U. \& Zwitter, T., 1997, A\&A, 318, 269
\bibitem[2010]{newton} Newton, J. \& Puckett, T., 2010, CBET, 2532, 1
\bibitem[2006]{patat06} Patat, F. \& Romaniello, M., 2006, PASP, 118, 146
\bibitem[2009]{patat09} Patat, F., Baade, D., H\"oflich, P., et al., 2009, A\&A, 508, 229
\bibitem[2010]{patat10} Patat, F., Maund, J.R., Benetti, S., et al., 2010,
\bibitem[1975]{serkowski} Serkowski, K., Matheson, D.S. \& Ford, V.L.,
	1975, ApJ, 196, 261
	\bibitem[1990]{schlegel90} Schlegel, E.M., 1990, MNRAS, 244, 269
\bibitem[1998]{schlegel} Schlegel, D.J., Finkbeiner, D.P. \& Davis, M.,
        1998, ApJ, 500, 525
\bibitem[1992]{schmidt} Schmidt, G.D., Elston, R., Lupie, O.L., 1992, 104, 1563
\bibitem[2010]{smith} Smith, N., et al., 2010, ApJ, submitted
  (arXiv:1011.4150v1)
\bibitem[2009]{trundle} Trundle, C., et al., 2009, A\&A, 504, 945
\bibitem[2001]{wang01} Wang, L., Howell, D.A., H\"oflich, P. \& Wheeler, J.C.,
        2001, ApJ, 550, 1030
\bibitem[2008]{wang08} Wang, L., \& Wheeler, J.C., 2008, ARAA, 46, 433
\bibitem[1992]{whittet} Whittet, D.C.B. et al., 1992, ApJ, 386, 562 
\end{thebibliography}
\end{document}